\documentclass[twocolumn,amsmath,nofootinbib,amssymb]{revtex4}

\usepackage{epsfig}
\usepackage{graphicx}
\usepackage{dcolumn}
\usepackage{bm}
\usepackage{color}

\def\thebiblio#1{
\begin{center}\bf \large References
\end{center}
\list
{[\arabic{enumi}]}{\settowidth\labelwidth{#1.}\leftmargin\labelwidth
 \advance\leftmargin\labelsep
 \usecounter{enumi}}
 \def\newblock{\hskip .11em plus .33em minus -.07em}
 \sloppy
 \sfcode`\.=1000\relax}

\begin{document}
\preprint{}
\title{Curvaton reheating allows TeV Hubble scale in NO inflation}

\author{J.C. Bueno S\'anchez}
 \email{j.buenosanchez@lancaster.ac.uk}
\author{K. Dimopoulos}%
 \email{k.dimopoulos1@lancaster.ac.uk}
\affiliation{%
Physics Department, Lancaster University\\}%

\date{\today}

\begin{abstract}
Curvaton reheating is studied in non-oscillatory (NO) models of inflation,
with the aim to obtain bounds on the parameters of curvaton models and find out
whether low scale inflation can be attained. Using a
minimal curvaton model, it is found that the allowed parameter space is
considerably larger than in the case of the usual oscillatory inflation models.
In particular, inflation with Hubble scale as low as 1~TeV is comfortably
allowed.
\end{abstract}


\maketitle


Inflation is to date the most compelling solution for the horizon and flatness
problems of big bang cosmology. Recently, observational data at high precision
have confirmed that structure formation and the CMB anisotropy are due
to the existence of superhorizon curvature perturbations, which need an acausal
mechanism to be created and, therefore, are further evidence of inflation.

The majority of inflationary models suggest that inflation took place
at energy comparable to that of grand unification. This is because, in most
inflationary models, the curvature perturbations are due to the quantum
fluctuations of the inflaton field; the field controlling the dynamics of
inflation. In this case the inflationary energy scale is \cite{book}
\begin{equation}
V_*^{1/4}=0.027\epsilon_*^{1/4}m_P\;,
\label{V*}
\end{equation}
where \mbox{$m_P=2.4\times 10^{18}\,$GeV} is the reduced Planck mass and
\mbox{$\epsilon\equiv-\dot H/H^2$} is one of the so-called slow-roll
parameters, where \mbox{$H\equiv\dot a/a$} with $a$ being the scale factor
and the dot denoting derivative with respect to time. In the above, `*' denotes
the epoch when the cosmological scales exit the horizon during inflation.
For inflation to occur one needs \mbox{$\epsilon<1$}, which is
equivalent to \mbox{$\ddot a>0$}. In slow-roll models of inflation
\mbox{$\epsilon\ll 1$}, which means \mbox{$H\simeq\,$constant}. However,
slow-roll inflation is not exactly de Sitter expansion, because it has to end
for the hot big bang to begin. Therefore, albeit small, in most models
$\epsilon$ is not tiny. In fact, it turns out that, typically,
\mbox{$\epsilon_*\sim 1/N_*$}, where $N_*$ is the remaining number of
inflation e-folds when the cosmological scales exit the horizon. $N_*$ ranges
between 40-65 for inflation scales between grand unification and Big
Bang Nucleosynthesis (BBN). Hence, from Eq.~(\ref{V*}) we see that
\mbox{$V_*^{1/4}\sim 10^{15-16}\,$GeV}. Exceptions to this rule exist, where
$\epsilon$ is exponentially suppressed due to appropriate features of the
scalar potential (e.g. see Ref.~\cite{Aterm}) but they are typically
fine-tuned (e.g. see Ref.~\cite{ours}).

However, in recent years, advances in string theory have turned the
attention at progressively lower energy scales. In particular, large
extra dimensions suggest that the fundamental scale may by much
smaller than $m_P$ and the scale of grand unification, possibly near
the electroweak scale instead (e.g. see \cite{TeV}). This means that
the hot big bang cannot be extended up to grand unification energies
and so inflation must take place at lower scale. Moreover, in the
search for the inflaton field, the most promising candidates are
either flat directions in supersymmetric theories or string moduli
fields, which correspond to the size and shape of the compactified
extra dimensions. However, in the MSSM and its extensions the
corresponding scales are typically much less than the grand unified
scale. Similarly, for string moduli one expects the scalar potential
to vary significantly (i.e. \mbox{$\delta V/V\sim 1$}) over
distances $\sim m_P$ in field space, which means that the typical
scale for modular inflation is
\mbox{$V_*^{1/4}\sim\sqrt{m_{3/2}m_P}\sim 10^{10.5}\,$GeV} (i.e.
\mbox{$H_*\sim 1\,$TeV}) for moduli with masses of order
\mbox{$m_{3/2}\sim 1\,$TeV}, such as string axions. Recently, many
possibilities for inflation in string theory have arisen in the
context of the string landscape \cite{cline}. For example,
metastable supersymmetric vacua \cite{meta} may account for the
vacuum density of inflation. This sets the value of $V_*$ at the
supersymmetry breaking scale which corresponds to the above
mentioned scale or lower. This is why there has been growing
interest in achieving inflation at low energy scales.

One way to liberate inflation from the constraint in Eq.~(\ref{V*}) is to
consider that the curvature perturbations generated by inflation are due
to the quantum fluctuations of a field {\em other} than the inflaton, in which
case Eq.~(\ref{V*}) turns into an upper bound \cite{liber}. This,
so called curvaton field, does not influence the dynamics of inflation but
becomes important after inflation has ended, when it imprints its
curvature perturbation onto the Universe \cite{curv}. Under this hypothesis, it
is possible to relax this constraint substantially when the curvaton model is
specially designed to do so \cite{low,laza}. However, generically, for a
minimal curvaton model, there is still a lower bound on the inflationary scale
of order \mbox{$V_*^{1/4}\gtrsim 10^{12}\,$GeV}
(i.e. \mbox{$H_*\gtrsim 10^6\,$GeV}) \cite{lyth}, when considering
inflation, which ends through oscillations of the inflaton field, as usually
assumed.

In this letter
we show that, for a minimal curvaton model, the inflation scale can be
substantially lower when considering a non-oscillatory (NO) model of inflation.
In such a model the inflaton corresponds to a flat direction with runaway
behaviour, typical, for example, of the scalar potential of the so-called
geometric string moduli fields. In such models after inflation ends the
Universe remains dominated by the inflaton field which is now fast-rolling down
its runaway scalar potential \cite{NO}. Since the inflaton does not oscillate
and does not decay after the end of inflation, reheating occurs by other
means.\footnote{NO models have been particularly useful for quintessential
inflation (e.g. see Ref.~\cite{eta,juan} and references therein) because the
inflaton field can survive until today and be responsible for the dark energy
at present.}
Recently, a curvaton field has been assumed to be responsible for reheating in
NO inflation models \cite{curvreh,eta,juan}.
However, up to now the parameter space for the inflationary scale
(in particular low-scale inflation) has not been studied.


We begin by outlining the particulars of NO inflation. In NO models, after the
end of inflation, the inflaton field becomes dominated by its kinetic density
\mbox{$\rho_\phi\sim\rho_{\rm kin}\equiv\frac12\dot\phi^2$} \cite{NO}.
The equation of motion for the field becomes
\mbox{$\ddot\phi+3H\dot\phi\simeq 0$}, which is oblivious to the scalar
potential and suggests that \mbox{$\rho_\phi\propto a^{-6}$}. This gives rise
to a period of so-called kination, where \mbox{$a\propto t^{1/3}$} \cite{kin}.
Because the dilution of the inflaton's density is so drastic, the Universe can
be reheated by the thermal bath due to the decay products of a curvaton field,
which also accounts for the density perturbations in the Universe
\cite{curvreh}. We should stress here that our treatment below is independent
of the NO inflation model since the scalar potential for the inflaton sector is
negligible during the evolution of the curvaton field.

We consider a minimal curvaton scenario where the curvaton field $\sigma$ is
characterised by the scalar potential
\begin{equation}
V(\sigma)=\frac12 m^2\sigma^2.
\label{Vs}
\end{equation}
In order to act as a curvaton $\sigma$ must undergo particle production during
inflation. Hence, it needs to be effectively massless. This implies the bound
\mbox{$m<H_*$}. An even tighter bound, however, is due to
spectral index considerations. Indeed , for the curvaton we have
\mbox{$n_s-1\simeq 2\eta=\frac23\left(m/H_*\right)^2$},
where \mbox{$\eta\equiv
m_P^2\frac{1}{V_*}\frac{\partial^2V}{\partial \sigma^2}$} \cite{curv}
with \mbox{$V_*\simeq 3H_*^2m_P^2$} being the density during inflation.
Since observations do not favour a blue spectrum, we require that $n_s$ is
at most \mbox{$n_s\simeq 1.00$}, which is still marginally acceptable.%
\footnote{Smaller values of $n_s$ can be attained if $\epsilon$ is not tiny,
in which case \mbox{$n_s-1\simeq 2(\eta-\epsilon)$} \cite{curv}.}
This means that $m$ must satisfy the bound
\begin{equation}
m\leq 0.1\,H_*\;.
\label{msbound}
\end{equation}

We assume that the inflaton's contribution to the curvature perturbation is
negligible, as is the norm under the curvaton hypothesis \cite{curv}. In any
case, it is difficult to obtain a sizable curvature perturbation in low scale
inflation \cite{liber} (See, however, Ref.~\cite{Aterm}).
Since the Universe is reheated by the decay products of the curvaton which
dominate the scalar field background, the observed curvature perturbation
\mbox{$\zeta=4.8\times 10^{-5}$} is entirely due to the curvaton, i.e.
\begin{eqnarray}
\zeta & = & \zeta_\sigma\equiv
-H\left.\frac{\delta\rho_\sigma}{\dot\rho_\sigma}\right|_{\rm dec}=
\frac13\left.\frac{\delta\rho_\sigma}{\rho_\sigma}\right|_{\rm dec}=
\frac23\left.\frac{\delta \sigma}{\sigma}\right|_{\rm dec}\nonumber\\
 & = & \frac23\left.\frac{\delta \sigma}{\sigma}\right|_{\rm osc}
\simeq\frac23\left.\frac{\delta \sigma}{\sigma}\right|_*\!=
\frac{H_*}{3\pi \sigma_*}\;\Rightarrow\;\sigma_*\sim\frac{H_*}{\zeta}\,,
\label{zs}
\end{eqnarray}
where we used that, before its decay (denoted as `dec'), the curvaton undergoes
quasi-harmonic oscillations in the scalar potential of Eq.~(\ref{Vs}), so that
\mbox{$\rho_\sigma=2\overline{V(\sigma)}\propto a^{-3}$} \cite{turner}, where
$\overline{V(\sigma)}$ is the average value of the potential density over many
oscillations. Since the oscillating curvaton satisfies the same linear equation
of motion as its oscillating perturbation,
(namely \mbox{$\ddot\sigma+3H\dot\sigma+m^2\sigma=0$})
the fractional perturbation $\delta\sigma/\sigma$ remains constant and equal to
its value at the onset of the oscillations, denoted as `osc'. Before the onset
of the oscillations, both the value of the field and its perturbation are
frozen so the fractional perturbation remains constant again and equal to its
value at horizon exit, denoted by `*'. As is the case with particle production
of scalar fields, the value of the curvaton perturbation at horizon exit is
determined by the Hawking temperature: \mbox{$\delta\sigma_*=H_*/2\pi$}. Note
that Eq.~(\ref{zs}) implies that \mbox{$\sigma_*\gg H_*$}, which guarantees
that the curvature perturbations are predominantly Gaussian \cite{curv}.

After the end of inflation, the inflaton becomes kinetically dominated,
while the curvaton remains frozen because \mbox{$m\ll H_*$} and its motion is
overdamped. This means that the density parameter of the
curvaton \mbox{$\Omega\equiv\rho_\sigma/\rho$} scales as
\mbox{$\Omega\propto a^6$} until $H(t)$ drops down to \mbox{$H\sim m$}, when
the curvaton unfreezes and begins its quasi-harmonic oscillations.
At this time we have
\begin{equation}
\Omega_{\rm osc}\sim\left(\frac{\sigma_*}{m_P}\right)^2\,.
\label{Oosc}
\end{equation}
After the onset of the oscillations, the curvaton density decreases as
\mbox{$\rho_\sigma\propto a^{-3}$} so that \mbox{$\Omega\propto a^3$}.

Now, there are two possibilities, depending on whether the curvaton decays
before its density dominates the Universe or afterwards. Let us assume firstly
that the curvaton decays before domination. In this case, using that
during kination \mbox{$a\propto H^{-1/3}$}, one readily obtains that the
curvaton density parameter at the time of curvaton decay is
\begin{equation}
\Omega_{\rm dec}\sim\frac{m}{\Gamma}\left(\frac{\sigma_*}{m_P}\right)^2,
\label{Odec}
\end{equation}
where $\Gamma$ is the curvaton decay rate. The decay products of the
curvaton redshift as relativistic matter \mbox{$\rho_\gamma\propto
a^{-4}$}, which means that their density parameter grows as
\mbox{$\Omega_\gamma\propto a^2$}. Reheating is achieved when this
density parameter reaches unity and the Universe becomes dominated
by the thermal bath of the curvaton decay products. Using that
\mbox{$H_{\rm dom}\sim T_{\rm reh}^2/m_P$}, it is easy to find
\begin{equation}
T_{\rm reh}\sim\sqrt{m_P\Gamma}\left(\frac{m}{\Gamma}\right)^{3/4}
\left(\frac{H_*}{\zeta m_P}\right)^{3/2},
\label{Treh}
\end{equation}
where we also employed Eq.~(\ref{zs}) and we denoted with `dom' the epoch when
the inflaton's density becomes subdominant. From the above and using also that
\mbox{$\Gamma\geq H_{\rm dom}$} and \mbox{$T_{\rm reh}>T_{\rm BBN}$}
we obtain the bound
\begin{equation}
\framebox{
$H_*>(10\zeta^2)^{1/3}(T_{\rm BBN}^2m_P)^{1/3}\sim 10\,{\rm GeV}$
}
\,,
\label{Hbound}
\end{equation}
where we also considered Eq.~(\ref{msbound}) and demanded that reheating occurs
before 
BBN; the latter corresponding to temperature
\mbox{$T_{\rm BBN}\sim 1\,$MeV}. Thus, we see that inflation with
\mbox{$H_*\sim 1\,$TeV} is possible to accommodate with a minimal curvaton
model.

Let us consider, now, the case when the curvaton decays after it dominates the
Universe. In this case the oscillating curvaton dominates when
\begin{equation}
H_{\rm dom}\sim m\left(\frac{\sigma_*}{m_P}\right)^2.
\label{Hdom}
\end{equation}
Reheating is, now, achieved at the decay of the curvaton, so that
\begin{equation}
T_{\rm reh}\sim\sqrt{m_P\Gamma}\,.
\label{Treh1}
\end{equation}
Using that \mbox{$\Gamma\leq H_{\rm dom}$} and also
\mbox{$T_{\rm reh}>T_{\rm BBN}$}, we arrive once more at the bound in
Eq.~(\ref{Hbound}), where we also considered Eq.~(\ref{msbound}).


Additional bounds on the parameters are obtained as follows. For the decay
rate of the curvaton we can write \mbox{$\Gamma=g^2m$}, where $g$ is the
coupling of the curvaton to its decay products. The expected range for this
coupling is:
\begin{equation}
\frac{m}{m_P}\lesssim g\lesssim 1\,,
\label{grange}
\end{equation}
where the lower bound is due to gravitational decay.

Assume at first that the curvaton decays before domination. Then, from
Eq.~(\ref{Treh}), we have \mbox{$T_{\rm reh}\propto\Gamma^{-1/4}$}.
Consequently, combining the lower bound in Eq.~(\ref{grange}) with the
requirement that \mbox{$T_{\rm reh}>T_{\rm BBN}$} and Eq.~(\ref{Treh}), we
obtain the bound
\begin{equation}
H_*\gtrsim\zeta(T_{\rm BBN}^2m_P)^{1/3}\sim 1\,{\rm GeV}\,,
\label{Hbound1}
\end{equation}
which is weaker than the bound in Eq.~(\ref{Hbound}). Also, using that
\mbox{$T_{\rm reh}>T_{\rm BBN}$}, Eq.~(\ref{Treh}) results in
\begin{equation}
g<\frac{m_Pm}{T_{\rm BBN}^2}\left(\frac{H_*}{\zeta m_P}\right)^3
\leq 10^{41}\zeta^{-3}\left(\frac{H_*}{m_P}\right)^4,
\label{gbound1}
\end{equation}
where, in the last inequality, we used Eq.~(\ref{msbound}). Comparing this
bound with the lower bound in Eq.~(\ref{grange}) it can be easily verified
(with the help of Eq.~(\ref{Hbound1})) that there is always parameter space
for $g$. Using also that \mbox{$g\lesssim 1\leq 0.1\,H_*/m$} according to
Eqs.~(\ref{msbound}) and (\ref{grange}), it can be shown that the bound in
Eq.~(\ref{gbound1}) is tighter that the upper bound in Eq.~(\ref{grange})
only for \mbox{$H_*< 10\,$TeV}. Finally, since in the case of decay before
domination we have \mbox{$\Gamma\geq H_{\rm dom}\sim T_{\rm reh}^2/m_P$},
employing Eq.~(\ref{Treh}) the upper bound in Eq.~(\ref{grange}) suggests
\begin{equation}
\sigma_*\lesssim m_P\Rightarrow H_*\lesssim\zeta m_P\sim 10^{14}\,{\rm GeV}\,,
\label{sbound}
\end{equation}
where we also used Eq.~(\ref{zs}).

Now, let us consider the case when the curvaton decays after domination,
where $T_{\rm reh}$ is given by Eq.~(\ref{Treh1}). Combining the requirement
that \mbox{$T_{\rm reh}>T_{\rm BBN}$} with the upper bound in
Eq.~(\ref{grange}) and using also Eq.~(\ref{msbound}), one finds the trivial
bound: \mbox{$H_*>10\,T_{\rm BBN}^2/m_P$}, which is way weaker than the bound
in Eq.~(\ref{Hbound}). The corresponding lower bound on $g$ is
\begin{equation}
g>\sqrt{10}\left(\frac{m_P}{H_*}\right)^{1/2}\frac{T_{\rm BBN}}{m_P}\,,
\label{gbound2}
\end{equation}
which is tighter than the lower bound in Eq.~(\ref{grange}) if
\mbox{$m<(T_{\rm BBN}^2m_P)^{1/3}$}. Finally, combining the
condition \mbox{$\Gamma<H_{\rm dom}$} with the lower bound in
Eq.~(\ref{grange}) one ends up again with the bound in
Eq.~(\ref{sbound}).

In summary, the allowed range for the inflationary Hubble scale is
\begin{equation}
10\,{\rm GeV}\leq H_*\leq 10^{14}\,{\rm GeV}\,,
\label{Hrange}
\end{equation}
while, for the decay coupling, the allowed range is
\begin{equation}
\!\,
\max\!\left\{\frac{T_{\rm BBN}}{\sqrt{m_Pm}},\frac{m}{m_P}\right\}
\!\lesssim g\lesssim\!
\min\!\left\{1, \frac{m_Pm}{T_{\rm BBN}^2}\!
\left(\frac{\sigma_*}{m_P}\right)^{\!3}\right\}\!\!.\!\!\!\!\!\!\!
\label{Grange}
\end{equation}

Another set of bounds is due to the possible overproduction of gravitational
waves (GWs) due to inflation. If inflation is followed by a phase whose
equation of state is stiffer than radiation, the spectrum of relic
gravitons features a spike (the slope grows with the frequency)
for modes re-entering the horizon during the stiff phase \cite{GW}.
Since in NO models there is a phase of kination right after inflation,
high frequency gravitons re-entering the horizon during kination
may disrupt BBN by increasing $H$. To avoid this, we require \cite{eta,GW}
\begin{equation}\label{gravBBN}
  I\equiv h^2\int_{\rm k_{\rm BBN}}^{k_*}\Omega_{\rm GW}(k)d\ln k\leq2\times
  10^{-6}\,,
\end{equation}
where $\Omega_{\rm GW}(k)$ is the GW density fraction with physical momentum
$k$ and $h=0.73$ is the Hubble constant $H_0$ in units of $100$ km/sec/Mpc.
Using the spectrum $\Omega_{\rm GW}(k)$ as computed in \cite{GW},
the above constraint can be written as \cite{eta}
\begin{equation}
I\simeq h^2\alpha_{\rm GW}\Omega_{\gamma}(k_0)\frac{1}{\pi^3}
  \frac{H_*^2}{m_P^2}\left(\frac{H_*}{H_{\rm
  dom}}\right)^{2/3}\,,
\end{equation}
where $\alpha_{\rm GW}\simeq 0.1$ is the GW generation efficiency
during inflation and \mbox{$\Omega_{\gamma}(k_0)=2.6\times 10^{-5}h^{-2}$}
is the density fraction of radiation at present on horizon scales.

If the curvaton field decays before domination, using
\mbox{$H_{\rm dom}\sim T^2_{\rm reh}/m_P$} and Eq.~(\ref{Treh}), the
bound in Eq.~(\ref{gravBBN}) becomes
\begin{equation}\label{gravBBN1}
\frac{H_*}{m}\left(\frac{\Gamma}{H_*}\right)^{1/3}\lesssim
\;24\,\zeta^{-2}\sim
10^{10}.
\end{equation}
If the curvaton decays after domination, using
Eqs.~(\ref{Hdom}) and (\ref{Treh1}), the
bound in Eq.~(\ref{gravBBN}) becomes
\begin{equation}\label{gravBBN2}
\frac{H_*}{m_P}\left(\frac{H_*}{\Gamma}\right)^{1/3}\lesssim 1\,.
\end{equation}
The above bounds may truncate further the ranges in
Eqs.~(\ref{Hrange}) and (\ref{Grange}).


Let us quantify the above with a couple of specific examples. Firstly, let us
choose \mbox{$H_*\sim 1\,$TeV} and \mbox{$m\sim 100\,$GeV} so that the bound
in Eq.~(\ref{msbound}) is saturated. In this case Eq.~(\ref{zs}) suggests that
\mbox{$\sigma_*\sim 10^8\,$GeV}. Using this, Eq.~(\ref{Grange}) becomes
\mbox{$10^{-13}\lesssim g\lesssim 10^{-4}$}. Using Eq.~(\ref{Hdom}) we find
\mbox{$H_{\rm dom}/\Gamma\sim 10^{-20}/g^2$}, which means that the curvaton
decays before domination if \mbox{$10^{-10}\lesssim g\lesssim 10^{-4}$},
whereas it decays after domination if
\mbox{$10^{-13}\lesssim g\lesssim 10^{-10}$}. In the former case
Eq.~(\ref{Treh}) gives \mbox{$T_{\rm reh}\sim g^{-1/2}10^{-5}\,$GeV}, while in
the latter case Eq.~(\ref{Treh1}) gives \mbox{$T_{\rm reh}\sim g10^{10}\,$GeV}.
Hence, the allowed range for the reheating temperature is:
\mbox{1 MeV$\,< T_{\rm reh}\lesssim\,$1 GeV}.
Since $\Gamma=g^2m$, it is straightforward to check that the bounds in
Eqs.~(\ref{gravBBN1}) and (\ref{gravBBN2}) are satisfied in the range shown in
Eq.~(\ref{Grange}).

Now, let us choose \mbox{$\sigma_*\sim m_P$} and \mbox{$m\sim 100\,$GeV},
corresponding to a string axion as curvaton. In this case Eq.~(\ref{zs})
suggests that \mbox{$H_*\sim 10^{14}\,$GeV}, which saturates the bound in
Eq.~(\ref{sbound}) and corresponds to inflation at the grand unified scale.
Using this, Eq.~(\ref{Grange}) becomes \mbox{$10^{-13}\lesssim g\lesssim 1$}.
Then, Eq.~(\ref{Hdom}) gives \mbox{$H_{\rm dom}/\Gamma\sim g^{-2}\gtrsim 1$}.
This means that the curvaton has to decay at or after domination. Thus, from
Eq.~(\ref{Treh1}) we have \mbox{$T_{\rm reh}\sim g10^{10}\,$GeV}, which
results in the range: \mbox{1 MeV$\,< T_{\rm reh}\lesssim\,10^{10}$ GeV}.
This time, however, the GW constraint turns out to be much tighter.
Because the curvaton decays at or after domination, we need to use
Eq.~(\ref{gravBBN2}). With the chosen values, the bound can only be
satisfied with \mbox{$g\sim 1$}, which saturates the upper bound in
Eq.~(\ref{grange}). Thus, the reheating temperature has to be
\mbox{$T_{\rm reh}\sim\,10^{10}$ GeV}, which seriously challenges
gravitino constraints.


In conclusion, we have studied in detail the parameter space for the inflation
scale in NO models with a minimal curvaton scenario being responsible for
reheating the Universe as well as for the curvature perturbations. We have
shown that low-scale inflation with $H_*$ as low as 10~GeV is feasible in this
case, regardless of whether the curvaton decays before or after dominating the
Universe. Our results are independent of the particular form of the NO
inflation model because, during the curvaton evolution, the inflaton sector is
oblivious of the scalar potential, since it is dominated by the kinetic density
of the inflaton.

\acknowledgements{
This work was supported (in part) by the European Union through the
Marie Curie Research and Training Network "UniverseNet"
(MRTN-CT-2006-035863) and by PPARC (PP/D000394/1).
}

\begin{thebiblio}{03}

{\small

\bibitem{book}
A.R.~Liddle and D.H.~Lyth,
{\em Cosmological Inflation and Large Scale Structure},
(Cambridge University Press, Cambridge U.K., 2000).

\bibitem{Aterm}
R.~Allahverdi, K.~Enqvist, J.~Garcia-Bellido and A.~Mazumdar,
Phys.\ Rev.\ Lett.\  {\bf 97} (2006) 191304;
R.~Allahverdi, K.~Enqvist, A.~Jokinen and A.~Mazumdar,
JCAP {\bf 0610} (2006) 007.

\bibitem{ours}
J.~C.~Bueno Sanchez, K.~Dimopoulos and D.~H.~Lyth,
JCAP {\bf 0701} (2007) 015.

\bibitem{TeV}
N.~Arkani-Hamed, S.~Dimopoulos and S.~Kachru,
hep-th/0501082.

\bibitem{cline}
J.~M.~Cline,
hep-th/0612129.

\bibitem{meta}
B.~Bergerhoff, M.~Lindner and M.~Weiser,
Phys.\ Lett.\  B {\bf 469} (1999) 61;
S.~A.~Abel, C.~S.~Chu, J.~Jaeckel and V.~V.~Khoze,
JHEP {\bf 0701} (2007) 089;
S.~A.~Abel, J.~Jaeckel and V.~V.~Khoze,
JHEP {\bf 0701} (2007) 015;
W.~Fischler, V.~Kaplunovsky, C.~Krishnan, L.~Mannelli and M.~A.~C.~Torres,
JHEP {\bf 0703} (2007) 107;
N.~J.~Craig, P.~J.~Fox and J.~G.~Wacker,
Phys.\ Rev.\  D {\bf 75} (2007) 085006.

\bibitem{liber}
K.~Dimopoulos and D.~H.~Lyth,
Phys.\ Rev.\ D {\bf 69}, 123509 (2004).

\bibitem{curv}
D.~H.~Lyth and D.~Wands,
Phys.\ Lett.\ B {\bf 524} (2002) 5;
T.~Moroi and T.~Takahashi,
Phys.\ Lett.\ B {\bf 522}, 215 (2001)
[Erratum-ibid.\ B {\bf 539}, 303 (2002)];
K.~Enqvist and M.~S.~Sloth,
Nucl.\ Phys.\ B {\bf 626} (2002) 395.

\bibitem{low}
K.~Dimopoulos, D.~H.~Lyth and Y.~Rodriguez,
JHEP {\bf 0502} (2005) 055;
M.~Postma,
JCAP {\bf 0405} (2004) 002.

\bibitem{laza}
K.~Dimopoulos and G.~Lazarides,
Phys.\ Rev.\  D {\bf 73} (2006) 023525;
K.~Dimopoulos,
Phys.\ Lett.\  B {\bf 634} (2006) 331

\bibitem{lyth}
D.~H.~Lyth,
Phys.\ Lett.\ B {\bf 579} (2004) 239.

\bibitem{NO}
G.~N.~Felder, L.~Kofman and A.~D.~Linde,
Phys.\ Rev.\  D {\bf 60} (1999) 103505.

\bibitem{eta}
K.~Dimopoulos,
Phys.\ Rev.\ D {\bf 68} (2003) 123506.

\bibitem{juan}
J.~C.~Bueno Sanchez and K.~Dimopoulos,
hep-th/0606223;
Phys.\ Lett.\  B {\bf 642} (2006) 294
[Erratum-ibid.\  B {\bf 647} (2007) 526].

\bibitem{curvreh}
B.~Feng and M.~z.~Li,
Phys.\ Lett.\ B {\bf 564} (2003) 169;
A.~R.~Liddle and L.~A.~Urena-Lopez,
Phys.\ Rev.\ D {\bf 68} (2003) 043517;
C.~Campuzano, S.~del Campo and R.~Herrera,
Phys.\ Lett.\ B {\bf 633} (2006) 149;
C.~Campuzano, S.~del Campo and R.~Herrera,
JCAP {\bf 0606} (2006) 017.

\bibitem{kin}
B.~Spokoiny,
Phys.\ Lett.\ B {\bf 315} (1993) 40;
M.~Joyce and T.~Prokopec,
Phys.\ Rev.\ D {\bf 57} (1998) 6022.

\bibitem{turner}
M.~S.~Turner,
Phys.\ Rev.\ D {\bf 28} (1983) 1243.

\bibitem{GW}
M.~Giovannini,
Phys.\ Rev.\ D {\bf 60} (1999) 123511;
V.~Sahni, M.~Sami and T.~Souradeep,
Phys.\ Rev.\ D {\bf 65} (2002) 023518.

}

\end{thebiblio}

\end{document}